\documentclass[final,twocolumn,twoside,times,5p]{elsarticle}
\biboptions{sort&compress}
\makeatletter
\def\ps@pprintTitle{%
 \let\@oddhead\@empty
 \let\@evenhead\@empty
 \let\@oddfoot\@empty
 \let\@evenfoot\@empty
}

\usepackage[utf8]{inputenc}
\usepackage[english]{babel}
\usepackage{amssymb,amsthm,amsmath,amstext,amsbsy,amsopn}
\usepackage{bbm}
\usepackage{graphicx}
\usepackage{isotope}
\usepackage{float}
\usepackage[normalem]{ulem}
\usepackage[dvipsnames]{xcolor}
\usepackage[T1]{fontenc}
\usepackage{subcaption}
\usepackage{lipsum}
\usepackage{gensymb}

\usepackage[numbers]{natbib}  
\usepackage[colorlinks=true,citecolor=blue,linkcolor=blue,urlcolor=blue]{hyperref}
\usepackage{orcidlink}

\newcommand{\ra}{\rangle}
\newcommand{\la}{\langle}

\begin{document}

\begin{frontmatter}



\title{Probing the shape evolution and shell structures in neutron-rich N=50 nuclei}

\address[ccs]{Center for Computational Sciences, University of Tsukuba, Tennodai, 
Tsukuba 305-8577, Japan}
\address[jaea]{Advanced Science Research Center, 
Japan Atomic Energy Agency, Tokai, 
Ibaraki 319-1195, Japan
}
\address[cns]{Center for Nuclear Study, The University of Tokyo, Hongo, Bunkyo-ku, 
Tokyo 113-0033, Japan}
           
\author[ccs]{Anil Kumar\orcidlink{0000-0003-3299-0435} }
\ead{anil@nucl.ph.tsukuba.ac.jp}
\author[ccs]{Noritaka Shimizu\orcidlink{0000-0003-1638-1066}}
\ead{shimizu@nucl.ph.tsukuba.ac.jp}
\author[ccs]{Takayuki Miyagi\orcidlink{0000-0002-6529-4164}}
\ead{miyagi@nucl.ph.tsukuba.ac.jp}
\author[ccs]{Yusuke Tsunoda\orcidlink{0000-0002-0398-1639}}
\ead{ytsunoda@nucl.ph.tsukuba.ac.jp}
\author[jaea,cns]{Yutaka Utsuno\orcidlink{0000-0001-5404-7320}}
\ead{utsuno.yutaka@jaea.go.jp}
             

\begin{abstract}
The structure of low-lying states of $N=50$ nuclei is investigated by the advanced Monte Carlo shell model (MCSM) in the $\pi{(fp)}$-$\nu{(sdg)}$ model space.   
We have employed the shell-model Hamiltonian based on the valence-space in-medium similarity renormalization group, with minimal phenomenological adjustments to the single-particle energies.   The MCSM results with the modified Hamiltonian nicely predict the shape coexistence of $^{78}$Ni, consistent with recent experimental data. The evolution of intrinsic shapes from the spherical shape to prolate shapes in the ground state of $N=50$ nuclei is discussed using the "T-plot" and effective single-particle energies, which visualize the intrinsic quadrupole deformation of the MCSM wave function.
The present result shows that the monopole part of the tensor force does not enhance the shape coexistence of $^{78}$Ni, unlike the case of $^{68}$Ni.
\end{abstract}


\end{frontmatter}




\section{Introduction}
\label{introduction}
Exploring the nuclear shell structure in unstable nuclei is one of the most interesting and not fully understood topics in nuclear physics, and it has become the forefront of both experimental and theoretical research in recent years. 
The shell structure of atomic nuclei can be visualized in terms of the single-particle orbits of protons and neutrons, similar to the orbits of electrons in an atom. Recent studies indicate that the shell structure is altered by the extreme ratio of proton and neutron numbers in the neutron-rich exotic nuclei, a phenomenon referred to as {\it shell evolution} \cite{Otsuka2020RMP, SORLIN2008602, NOWACKI2021103866}. Two types of shell evolution have been identified in previous studies \cite{Tsunoda2014RPRC, Otsuka_2016}, namely {\it type I} and {\it type II}. When the shell structure evolves along an isotopic or isotonic chain, it is referred to as {\it type I}. On the other hand, depending on the quantum numbers of the Hamiltonian eigenstate, shell structure can change within the same nucleus due to differences in nucleon occupancies, exhibiting collective effects such as deformation and weakening of the traditional shell closures, denoted as {\it type II}.  Due to shell evolution, exotic nuclei sometimes exhibit the disappearance of the standard magic number and the emergence of new magic numbers in some instances \cite{warner2004Nat, Otsuka2006PRL, Robert2009Nat, Steppenbeck2013Nat}, thereby leading to a peculiar manifestation of nuclear dynamics. For example, the traditional neutron magic  $N= 8, ~20, ~28, ~{\rm and} ~40$ numbers disappear, and new magic numbers like $N=14, ~16, ~32,~ {\rm and}~34$ emerge in the neutron-rich regions \cite{Becheva2006PRL, wimmer2010PRL, Steppenbeck2013Nat, Rogers2015PRC}.

Around $N=40$, a group of nuclei close to semi-magic or even doubly magic nuclei form the second Island of Inversion (IoI),  where deformation occurs instead of the expected spherical shape in their ground states \cite{Hannawald199PRL, Gade2010PRC, Rother2011PRL, Crawford2013PRL, Lenzi2010PRC}. The same nucleus exhibits these types of multiple nuclear shapes at similar energies, referred to as shape coexistence \cite{Heyde2011RMP, GARRETT2022103931, Otsuka_2016}.  Recently, many experimental and theoretical findings 
indicated the emergence of shape coexistence in neutron-rich nuclei $^{68}$Ni in the $N=40$ sub-shell closure  \cite{Sorlin2002PRL,  Dijon2012PRC, Suchyta2014PRC, Gaudefroy2009PRC, Lenzi2010PRC, Tsunoda2014RPRC}. The observed well-deformed band structure at low excitation energies becomes the ground state in Cr and Fe isotopes of $N=40$ nuclei, resulting in collective behavior, as seen in $^{64}$Cr and $^{66}$Fe \cite{Crawford2013PRL, Gade2010PRC, Ljungvall2010PRC, CORTES2020PLB, Rother2011PRL}. Further experimental measurements on Fe isotope in  $N=40$ IoI reveal that $^{70,72}$Fe still exhibits a collective nature,  which indicates the possible extension of IoI towards $N=50$ \cite{Santamaria2015PRL}. 

Recently, Taniuchi {\it et al.} \cite{Tanuuchi2019nature} provided the first experimental spectroscopy study of $^{78}$Ni, and the signature of shape coexistence was already identified from the neighboring nuclei in many recent experimental measurements \cite{yang2016PRL, Gottardo2016PRL, Nies2023PRL}. The $N=50$ isotone undergoes a sudden change in shape, posing a significant challenge for theoretical models. Various theoretical approaches have been performed, ranging from the shell model with different phenomenological \cite{Nowacki2016PRL, JGLI2023PLB} and ones based on chiral effective field theory to (beyond) mean-field calculations to explore the $N=50$ shell closure \cite{ TICHAI2024PLB, HU2024PLB}.  The large-scale shell model calculations predicted the $N=50$ IoI with deformed ground states rather than spherical in $^{76}$Fe and $^{74}$Cr \cite{Nowacki2016PRL, JGLI2023PLB}. More recently, the first $ab~initio$ coupled cluster calculations also predicted how the deformed configuration evolves in the ground states of $N=50$ nuclei from $^{78}$Ni to $^{70}$Ca \cite{HU2024PLB}. Tichai {\it et al.} \cite{TICHAI2024PLB} proposed the merging of the ab initio approach valence-space in-medium similarity renormalization-group (VS-IMSRG) with the density matrix renormalization group (DMRG) as a scalable alternative to interpreting the low-lying spectroscopy of $N=50$ isotones. However, in \cite{Tanuuchi2019nature}, the VS-IMSRG(2) calculations predicted a too large second $2^+$ excitation energy of $^{78}$Ni as compared to the experimental one. The large deviation may be obtained since all operators are truncated at the two-body level. Therefore, in this work, we attempt to modify the one-body part of the VS-IMSRG Hamiltonian phenomenologically as a possible option to capture the missing physics.

In this work, we construct a shell-model Hamiltonian in the $\pi(fp)$-$\nu(sdg)$ model space on {\it ab initio} based approach VS-IMSRG \cite{Stroberg2019ARNP} that is suitable for interpreting the nuclear structure of $N\sim 50$.  More recently, Purcell \textit{et al.} \cite{Purcell2025PRC} developed an empirical shell model Hamiltonian that considers $^{78}$Ni as core, using the VS-IMSRG Hamiltonian as a starting point. They employed two VS-IMSRG approaches:  VS-IMSRG(2) \cite{Stroberg2019ARNP} and VS-IMSRG($3{\rm f}_2$) \cite{BCHE2024PRC}, which introduces a new correction by including intermediate three-body operators through doubly nested commutators. Ref.~\cite{Purcell2025PRC} demonstrates that the VS-IMSRG($3{\rm f}_2$) approach provides a better input than IMSRG(2), which requires less modification to reproduce experimental data. 
In the present work, we also confirmed that the VS-IMSRG($3{\rm f}_2$) is a better starting point than VS-IMSRG(2) by tuning only single-particle energies (SPEs).
Therefore, we have adopted the VS-IMSRG(3${\rm f}_2$) Hamiltonian as a starting point and implemented a phenomenological correction on the one-body part of the Hamiltonian in order to reproduce the experimental data in $N=50$  region.

With the development of the radioactive ion-beam facilities of the new generation, several experiments were performed to investigate the neutron-rich $N=50$ shell closure \cite{yang2016PRL, Gottardo2016PRL, Nies2023PRL, Olivier2017PRL, welker2017PRL, VAJTA2018PLB, ORLANDI2015PLB}. For example, the low-lying states in $^{79}$Cu \cite{Olivier2017PRL} and $^{79}$Zn \cite{Nies2023PRL} nuclei near $^{78}$Ni have been established experimentally.
Using the  VS-IMSRG($3{\rm f}_2$)-based Hamiltonian,  we introduced minimal phenomenological corrections only to the SPEs to reasonably reproduce the experimental data of $^{79}$Cu (odd-even) and  $^{79}$Zn (even-odd) nuclei.
The neutron-rich $N=50$ region is both very demanding from the shell model perspective and computationally challenging. We have performed the large-scale shell model calculations by using the state-of-the-art advanced Monte Carlo Shell Model (MCSM) \cite{shimizu2017ps,shimizu2010PRC, shimizu2012ptep, OTSUKA2001PPNP}, since the large-scale MCSM calculations play a vital role in deepening our understanding of exotic, particularly in the context of ``shell evolution'', which has been strongly supported by the success of MCSM in the IoI \cite{Mizusaki1999PRC, Utsuno1999PRC, Tsunoda2014RPRC}.

This paper is organized as follows: the theoretical framework used in this work is discussed in Sec. \ref{sec:method} and the results and discussion are provided in Sec. \ref{sec:results}. Finally, Sec. \ref{sec:conclusion} summarizes the key features of this study.

\section{Model Space and Effective Hamiltonian}
\label{sec:method}
This section discusses the model space, effective Hamiltonian, and MCSM calculations used in the present study.  In the present work, we have carried out the shell-model calculations with core, in which the intrinsic shell-model Hamiltonian is defined as 

 \begin{equation}
     H = \sum_{ij} t_{ij} \hat{c}_i^\dagger\hat{c}_j + \sum_{i< j,k< l} {v}_{ijkl}\hat{c}_i^\dagger\hat{c}_j^\dagger\hat{c}_l\hat{c}_k,
 \end{equation}
 with the nucleon creation ($\hat{c}^\dagger$) and  annihilation ($\hat{c}$) operators. The notations $i$, $j$, $k$, and $l$ are specified for the single-particle states. The one- and two-body matrix elements (TBMEs) are denoted $t_{ij}$ and ${v}_{ijkl}$. 

The model space above the $^{60}$Ca inner core is used in the full $fp$ shell $\{0f_{7/2,5/2},~1p_{3/2,1/2}\}$ for protons and the full $sdg$ shell $\{0g_{9/2,7/2},~1d_{5/2,3/2},~2s_{1/2}\}$ for neutrons. 
Starting from the 1.8/2.0 (EM) nucleon-nucleon and three-nucleon interactions~\cite{Hebeler2011} expressed in the 13 major-shell harmonic-oscillator space with the frequency of 16 MeV, we decouple the above-mentioned valence space from its complement using the VS-IMSRG($3{\rm f}_{2}$) approximation~\cite{BCHE2024PRC} with the spherical Hartree-Fock calculated for $^{78}$Ni.
In this work, we use the TBMEs from the IMSRG($3{\rm f}_{2}$) without any phenomenological corrections. To make TBMEs globally applicable in the adopted valence space, however, we employ the following mass-dependent TBMEs:
\begin{equation}
{v}_{ijkl}(A)= \left(\frac{A}{78}\right)^{p}{v}_{ijkl}(A = 78).
\end{equation}
It turns out that $p=-0.3$ is the most consistent choice with the nucleus-dependent VS-IMSRG($3{\rm f}_{2}$) TBMEs and is the same value as found in the earlier works~\cite{Magilligan2021PhysRevC.104.L051302, Purcell2025PRC}.
Further, we have made phenomenological corrections in the SPEs. As discussed in the previous section, the new data for nuclei $^{79}$Cu \cite{Olivier2017PRL}, $^{79}$Zn \cite{Nies2023PRL}  near to $^{78}$Ni are good candidates for describing the proton and neutron single-particle states.  The proton SPEs of $\pi(1p_{1/2,3/2}, 0f_{5/2})$ orbitals are optimized to reproduce experimental data of low-lying energy levels $5/2^-$, $3/2^-$, and $1/2^-$ of $^{79}$Cu ($Z=29, ~N=50$) \cite{Olivier2017PRL} and the neutron SPEs of $\nu(0g_{9/2,7/2}, 1d_{5/2,3/2}, 2s_{1/2})$ orbital are optimized to reproduce the relative location of low-lying states $9/2^+$, $1/2^+$, and $5/2^+$ of the $^{79}$Zn ($Z=30,~ N=49$) \cite{Nies2023PRL}  reasonably. 
The size of the $Z=28$ proton gap is determined to reproduce the excitation energies of the first $2^+$ of Zn, Ge, and Se isotopes around the $N=50$. In practice, for the $^{79}$Cu and $^{79}$Zn isotopes, the root-mean-square deviation between theoretical and experimental energies was approximately 500 keV when using the original IMSRG SPEs. This deviation was significantly reduced to about 110 keV after the correction of the SPEs.

To perform the large-scale shall model calculations, we have used the advanced MCSM \cite{shimizu2010PRC, shimizu2012ptep, OTSUKA2001PPNP}. The advantage of the MCSM is that it allows the inclusion of many single-particle states, enabling the description of states with diverse nuclear structure behavior. 
MCSM calculation can describe spherical yrast states, deformed rotational bands, and nearly superdeformed bands using the same Hamiltonian within the same model space. Typical examples are $^{56}$Ni \cite{Mizusaki1999PRC, OTSUKA2001PPNP} and $^{68}$Ni \cite{Tsunoda2014RPRC}, where the MCSM calculations provide a nice description of those kinds of states. 
The nuclear wave function $|\Psi\rangle$  in MCSM is written as a superposition of the angular momentum and parity projected deformed Slater determinants (MCSM basis vectors) as
\begin{equation}\label{eqn:mcsmwave}
    |\Psi\rangle = \sum_{n=1}^{N_b}\sum_{K=-J}^{J}f_{nK}P^{J\pi}_{MK}|\phi_n\rangle,
\end{equation}
where $P_{MK}^{J\pi}$ is the angular-momentum and parity projector, $N_b$ is the number of Slater-determinant basis states, $f_{nK}$ denotes as amplitude, and $|\phi_n\rangle$ is a Slater determinants which can be defined as
\begin{equation}
|\phi_n\rangle =   \prod_\alpha\left(\sum_k D_{k\alpha}^{(n)}c_k^{\dagger}\right) |-\rangle,
\end{equation}
where $|-\rangle$ is an inert core.
$D_{k\alpha}^{(n)}$ is the amplitude determined so that the resultant energy is minimized.
Throughout this work, we use $N_{b}=100$, which is sufficiently large to obtain the converged results~\cite{shimizu2017ps}.

\section{Results and Discussions}\label{sec:results}
Our primary interest is the interpretation of the nuclear structure properties of nuclei near or on the neutron magic number $N=50$.  The advanced MCSM \cite{shimizu2010PRC, shimizu2012ptep, OTSUKA2001PPNP, shimizu2017ps}  calculations are carried out to calculate the nuclear structure properties of even $Z = 20 - 30$ by utilizing the modified Hamiltonian based on the VS-IMSRG($3{\rm f}_2$). Hereafter, the calculated results will be referred to as ``MCSM'' in the entire manuscript.

\begin{figure}[t]
\centering
\includegraphics[width=\columnwidth]{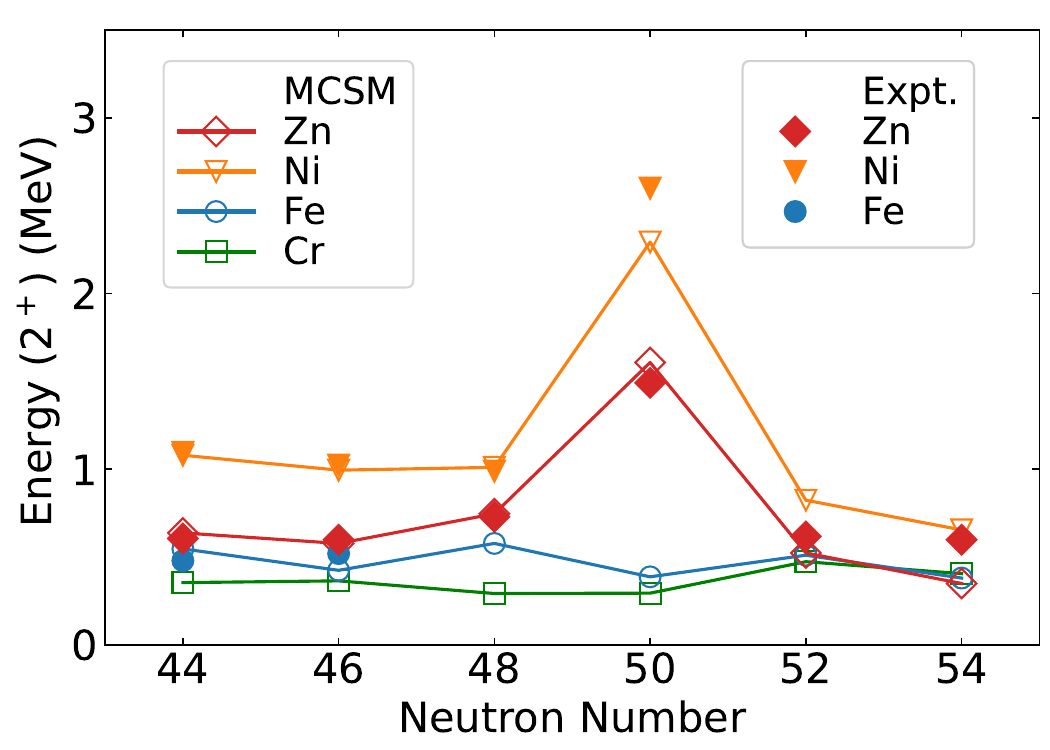}
\caption{ First $2^+$ excitation energies of the Cr, Fe, Ni, and Zn isotope chains with $N= 44-54$ compared with the available experimental data \cite{nndc2024}.
}\label{fig:n50_2p_energy}
\end{figure}

\begin{figure*}[h]
\centering
 \hspace{-1.02cm}
\includegraphics[width=18.0cm, height = 9.0cm]{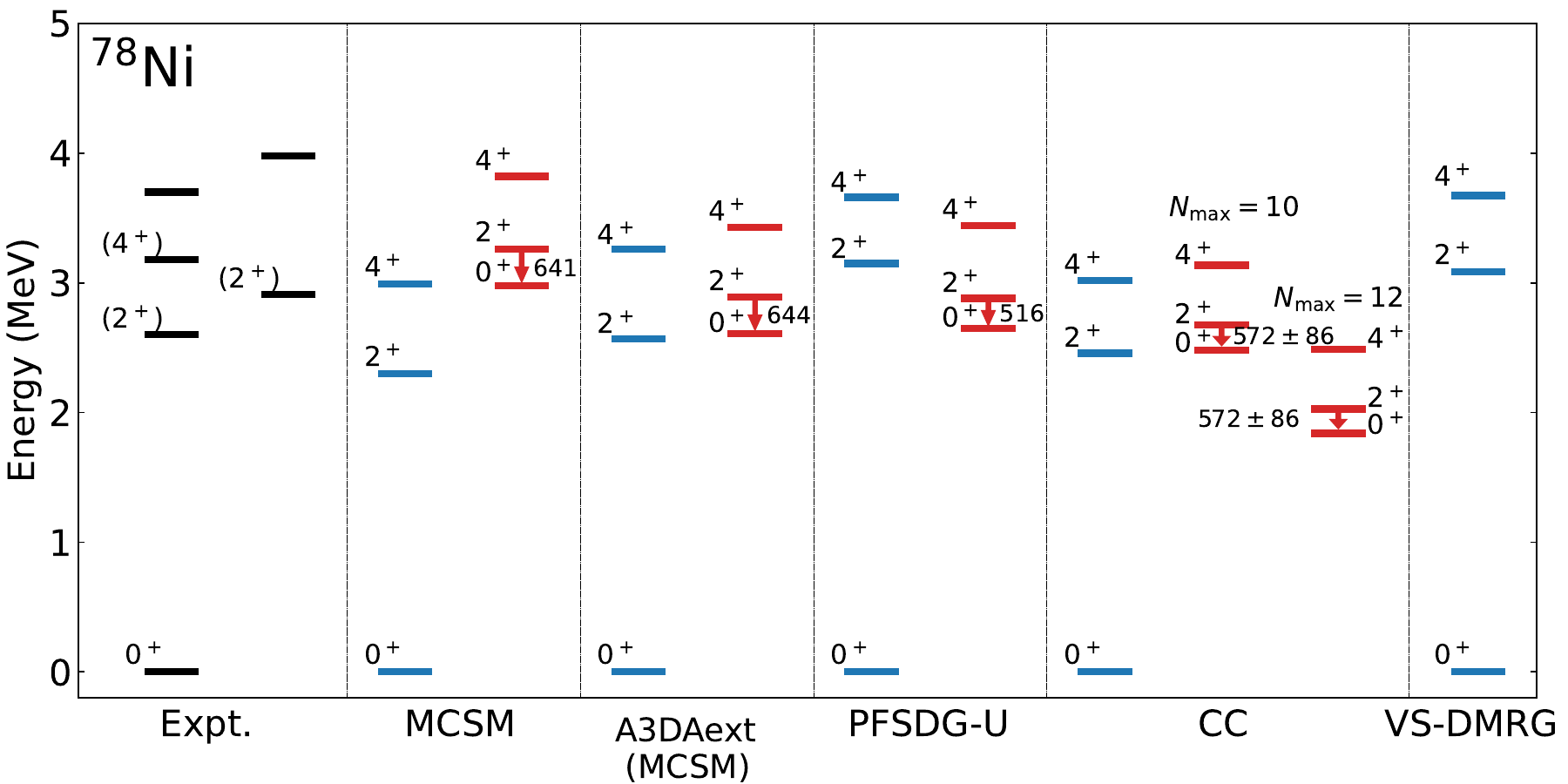}
\caption{Energy levels of $^{78}$Ni obtained from the present MCSM calculations (labeled with MCSM) in comparison to the available experimental data \cite{Tanuuchi2019nature} and several theoretical predictions such as the MCSM calculations with the extended version of A3DA Hamiltonian (labeled with A3DAext) \cite{Tanuuchi2019nature}, large-scale shell model calculation with PFSDG-U interaction (labeled with PFSDG-U) \cite{Nowacki2016PRL},  and two more recent \textit{ab initio} calculations, namely the coupled-cluster (CC) calculations \cite{HU2024PLB} and the valence-space density matrix renormalization group (VS-DMRG) \cite{TICHAI2024PLB}. The $B(E2; 2^+\to 0_2^+)$ values are depicted by arrow in ${\rm e}^2{\rm fm}^4$ units.   }\label{fig:spec_78Ni}
\end{figure*}

Figure \ref{fig:n50_2p_energy} shows the MCSM results of the first $2^+$ excitation energy of the Cr, Fe, Ni, and Zn isotopes as a function of neutron number $N=44-54$ compared with the available experimental data \cite{nndc2024}. The $Ex(2^+)$ values obtained by the MCSM are in good quantitative agreement with the experimental data, while that of $^{78}$Ni shows a slight underestimation. 
According to the MCSM results, the Ni isotope exhibits a peak, indicating the $N=50$ closed shell structure, which is consistent with the experimental data. However, such a peak structure disappears in the Cr and Fe isotopes, and these $2^+$ energies exhibit an almost flat and stable trend across the isotopic chains, indicating the signature of the entrance to the {\it island of inversion} and the beginning of deformation. Similar behavior in $N=50$ has also been discussed in the previous shell-model studies \cite{Nowacki2016PRL, JGLI2023PLB}. Moreover, our MCSM calculations yield the lowest $2^+$ energy for $^{74}$Cr, indicating maximal deformation at the mid-proton shell.  This point will be discussed later through the "T-Plot" analysis.

We show in Fig.~\ref{fig:spec_78Ni} the MCSM energy levels of $^{78}$Ni in comparison to the available experimental data of Taniuchi \textit{et al.} \cite{Tanuuchi2019nature} and various theoretical predictions, such as the MCSM results based on the extended version of A3DA Hamiltonian (labeled with A3DAext) \cite{Tanuuchi2019nature}, the large-scale shell model calculations with the PFSDG-U interaction of Nowacki \textit{et al.} \cite{Nowacki2016PRL}, and with more recent two first-principle calculations, namely the valence-space density matrix renormalization group (VS-DMRG)  \cite{TICHAI2024PLB} and the coupled-cluster (CC) calculations \cite{HU2024PLB}. In both of these {\it ab initio} calculations, the same Hamiltonian 1.8/2.0 (EM)~\cite{Hebeler2011} was used, which is also the starting point in the present study.
In Ref.~\cite{Tanuuchi2019nature}, the VS-IMSRG calculations predicted the second $2^+$ state at a high energy of 5.08 MeV, while it was experimentally observed at 2.91 MeV. 
With the modified Hamiltonian in the present work, the MCSM result nicely reproduces it within about 250 keV.
Moreover, we also observed two distinct bands in $^{78}$Ni: one particularly exhibits the traditional spherical shape associated with shell closure, while another displayed a pronounced deformed
shape at higher excitation energy. The nature of this deformed band will be further discussed in terms of "T-plots". The MCSM ground-state band is consistent with the experimental data and previous theoretical predictions. However, the $0^+_2$ state, which is expected to be the head of the deformed band, has not yet been observed experimentally. 
The present result predicts the head of the deformed band $0^+_2$ state with a slightly higher excitation energy compared to the previous predictions. 
Our MCSM predicted ratio $R_{4/2}\equiv E(4^+)/E(2^+)$ for this rotational band is 2.95, yielding a consistent value with the previous 3.43 for large-scale shell model \cite{Nowacki2016PRL}, 2.93 for MCSM with A3DAext interaction \cite{Tanuuchi2019nature}, and 3.42 for CC \cite{HU2024PLB}. The obtained  $R_{4/2}$ values from all theoretical predictions are comparable to the rigid rotor limit $R_{4/2}=3.33$.

\begin{figure}[!h]
\centering
\includegraphics[width=\columnwidth]{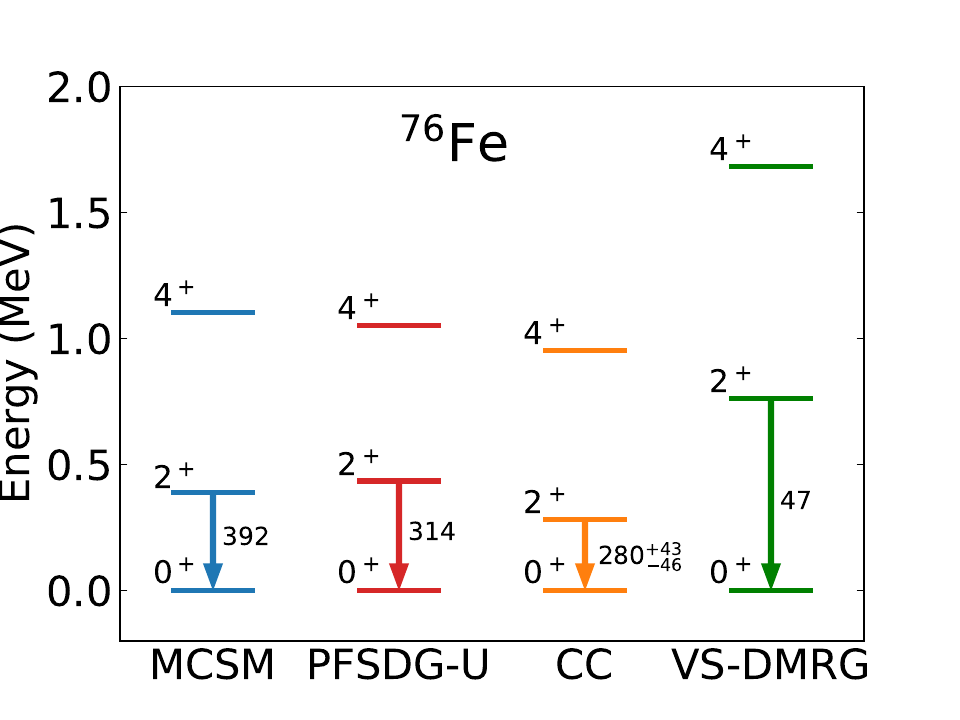}
\includegraphics[width=\columnwidth]{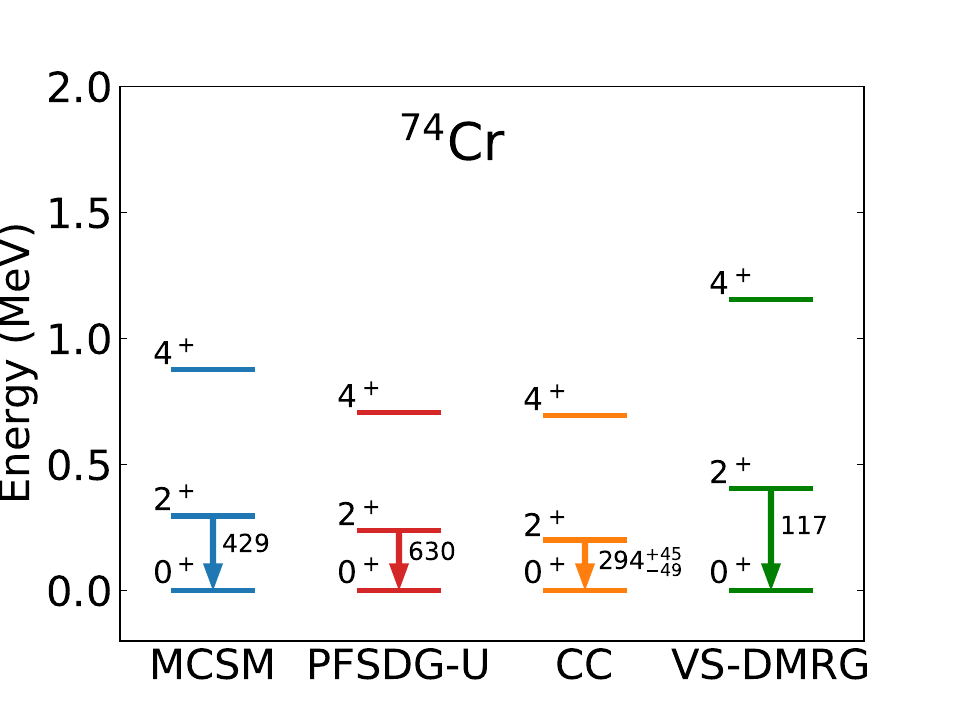}
\includegraphics[width=\columnwidth]{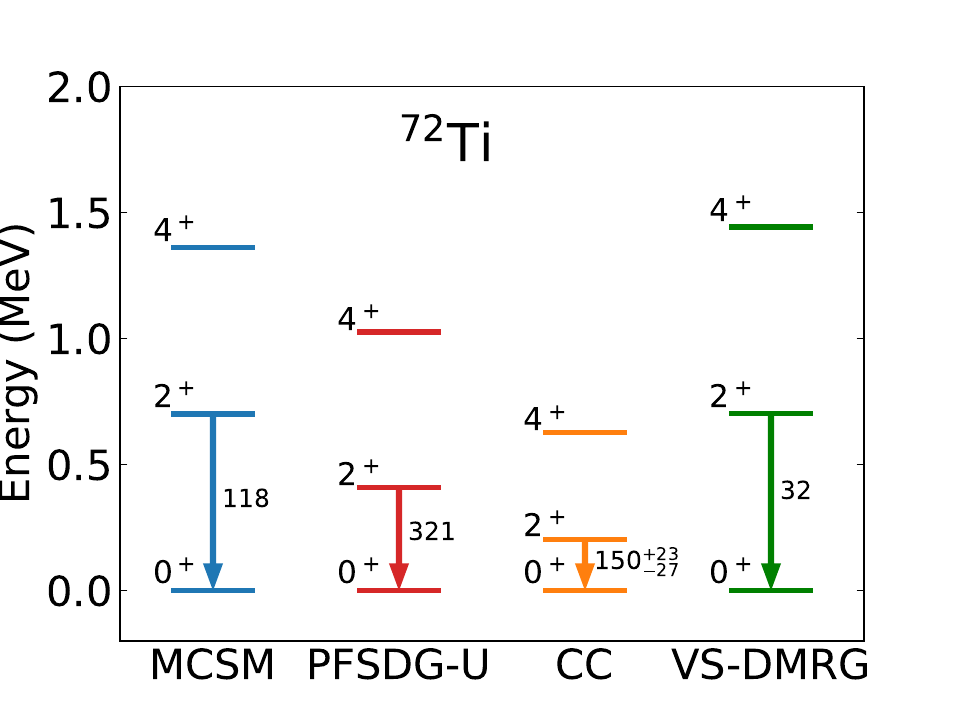}
\caption{ The MCSM excitation energy spectra of the $^{76}$Fe, $^{74}$Cr, and $^{72}$Ti with the modified VS-IMSRG($3{\rm f}_2$) in comparison to large-scale shell models predicted with PFSDG-U interaction \cite{Nowacki2016PRL}, coupled-cluster (CC) calculations \cite{HU2024PLB}, and valence-space density matrix renormalization group (VS-DMRG) \cite{TICHAI2024PLB}. The $B(E2;2^+_1\to 0_1^+)$ values are depicted by the arrow in ${\rm e}^2{\rm fm^4}$ units. 
}\label{fig:n50_spectra}
\end{figure}

The MCSM prediction of low-lying energy spectra of proton deficient $N=50$ nuclei, namely $^{76}$Fe, $^{74}$Cr, and $^{72}$Ti are presented in Fig.~\ref{fig:n50_spectra}. The obtained excitation spectra of these nuclei are consistent with the previous theoretical predictions. 
The previous shell-model studies \cite{Nowacki2016PRL, JGLI2023PLB} and ab initio coupled-cluster calculation \cite{HU2024PLB} have proposed the new IoI at $N=50$ near  $^{74}$Cr, below $^{78}$Ni. The present MCSM study also observes prolate states with larger deformation in $^{76}$Fe, $^{74}$Cr, and $^{72}$Ti, which arise from intruder configurations formed by neutron excitations across $N=50$. These states, shown in Fig. \ref{fig:n50_spectra}, exhibit similarities to previous theoretical predictions.

 We have also verified the rotational band structure by the $B(E2)$ values. In this work, we have phenomenologically modified the SPEs in the VS-IMSRG($3{\rm f}_2$) Hamiltonian. Therefore, the $B(E2)$ values are computed by using the effective charges $(e_\pi,e_\nu)=(1.31, 0.46)e$, as adopted in Ref.~\cite{Nowacki2016PRL}.  The obtained $B(E2)$ values have been shown in Figs. \ref{fig:spec_78Ni} and \ref{fig:n50_spectra}. In the case of $^{78}$Ni, the present MCSM calculation predicted the large $B(E2; 2_2^+\to0_2^+)$ = 641 $e^2{\rm fm}^4$ corresponding to the deformed band, which is consistent with the previous theoretical works.  In the case of $^{74}$Cr, the MCSM predicts the lowest $2^+$ state, which also corresponds to a large $B(E2)$ value.

\begin{figure*}[h]
\centering
\hspace{-0.68cm}
\vspace{0.3cm}
 \begin{subfigure}{0.20\textwidth}
 \phantomcaption\label{fig:pes_n50_a}
\includegraphics[width=4.1cm, height = 2.75cm]{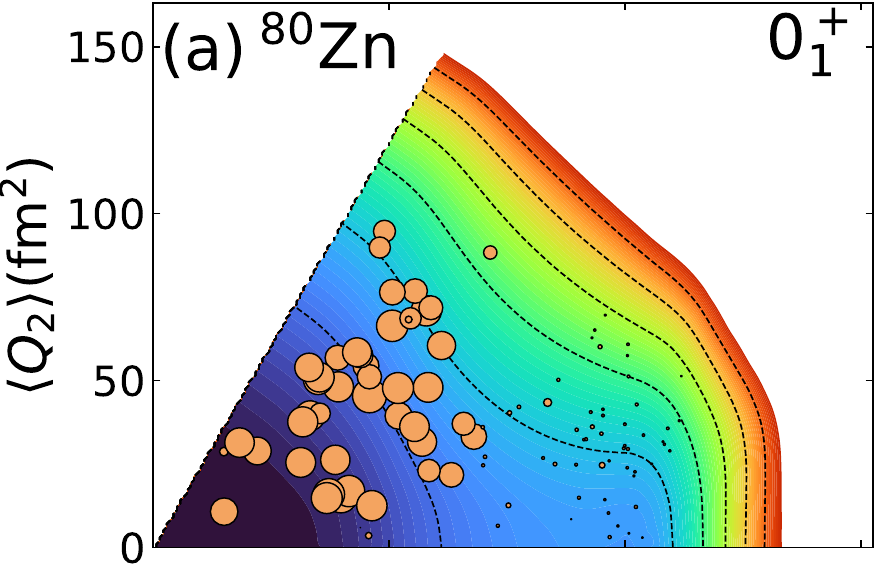}
\end{subfigure}
\hspace{5mm}
 \begin{subfigure}{0.20\textwidth}
 \phantomcaption\label{fig:pes_n50_b}
\includegraphics[width=3.5cm, height = 2.75cm]{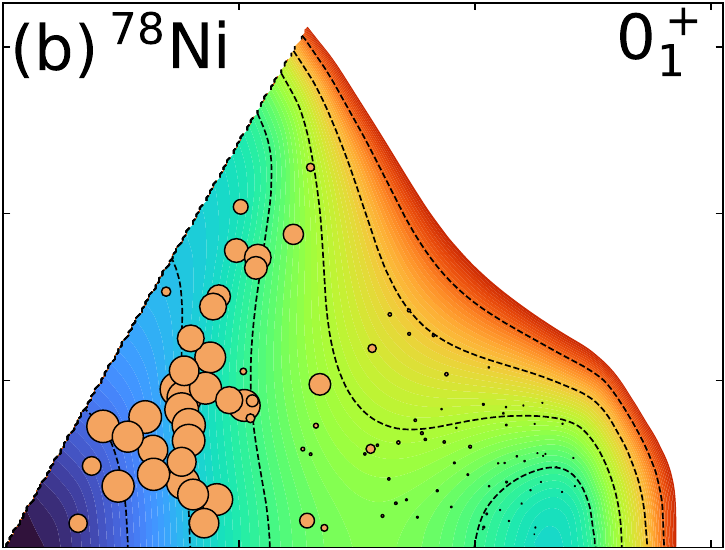}
\end{subfigure}
\hspace{-2mm}
 \begin{subfigure}{0.20\textwidth}
 \phantomcaption\label{fig:pes_n50_c}
\includegraphics[width=3.5cm, height = 2.75cm]{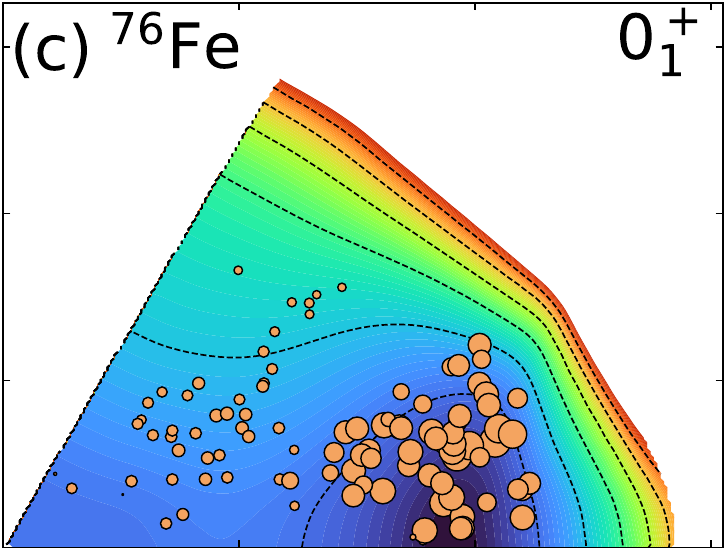}
\end{subfigure}
\hspace{-2mm}
\begin{subfigure}{0.20\textwidth}
\phantomcaption\label{fig:pes_n50_d}
\includegraphics[width=3.5cm, height = 2.75cm]{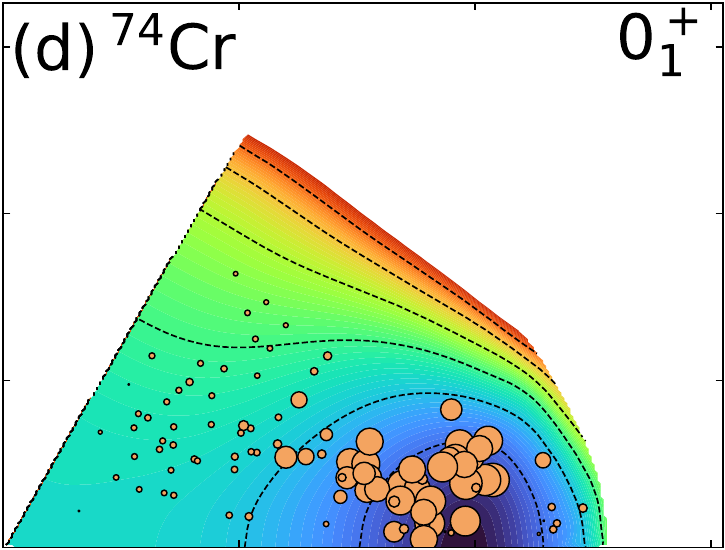}
\end{subfigure}
\hspace{-2mm}
 \begin{subfigure}{0.20\textwidth}
 \phantomcaption\label{fig:pes_n50_e}
\includegraphics[width=3.5cm, height = 2.75cm]{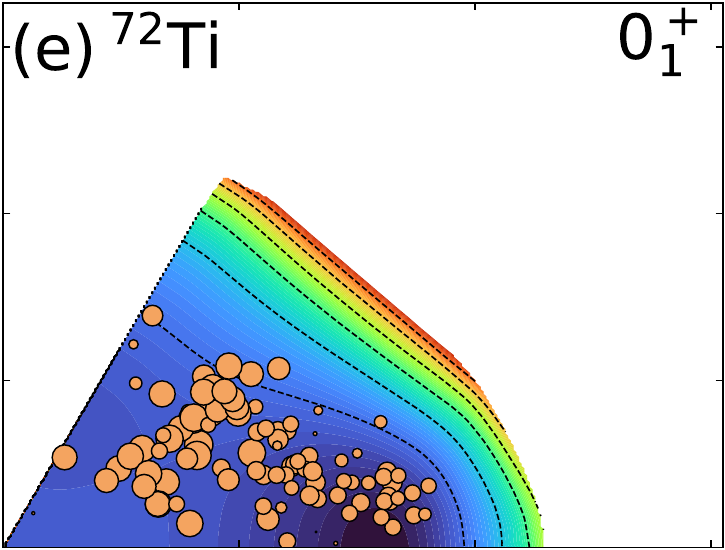}
\end{subfigure}
  \par\medskip 
  \vspace{-0.45cm}
\hspace{-0.66cm}
\begin{subfigure}{0.20\textwidth}
\phantomcaption\label{fig:pes_n50_f}
\includegraphics[width=4.2cm, height = 3.1cm]{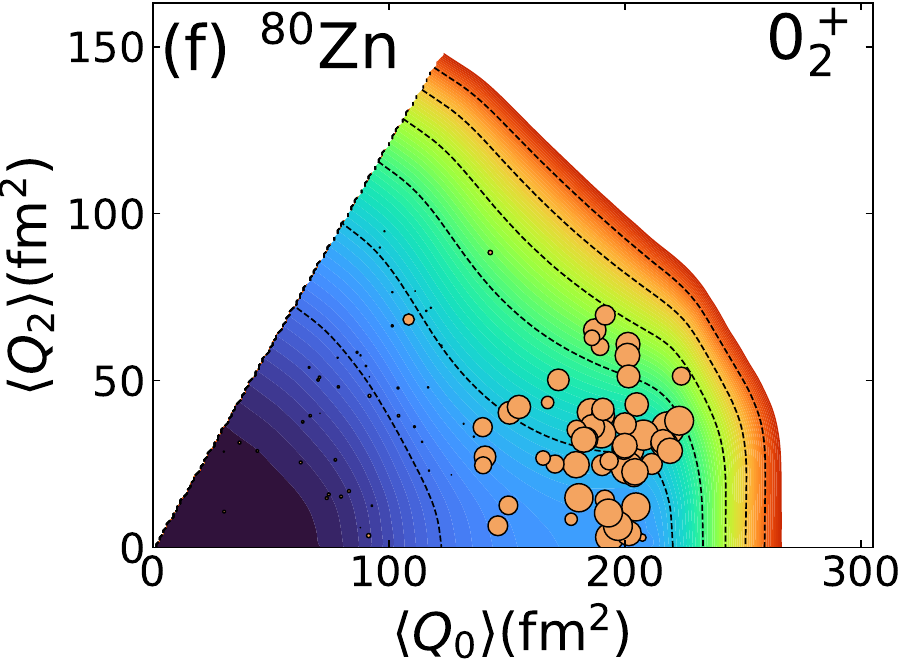}
\end{subfigure}
\hspace{5mm}
\begin{subfigure}{0.20\textwidth}
\phantomcaption\label{fig:pes_n50_g}
\includegraphics[width=3.6cm, height = 3.1cm]{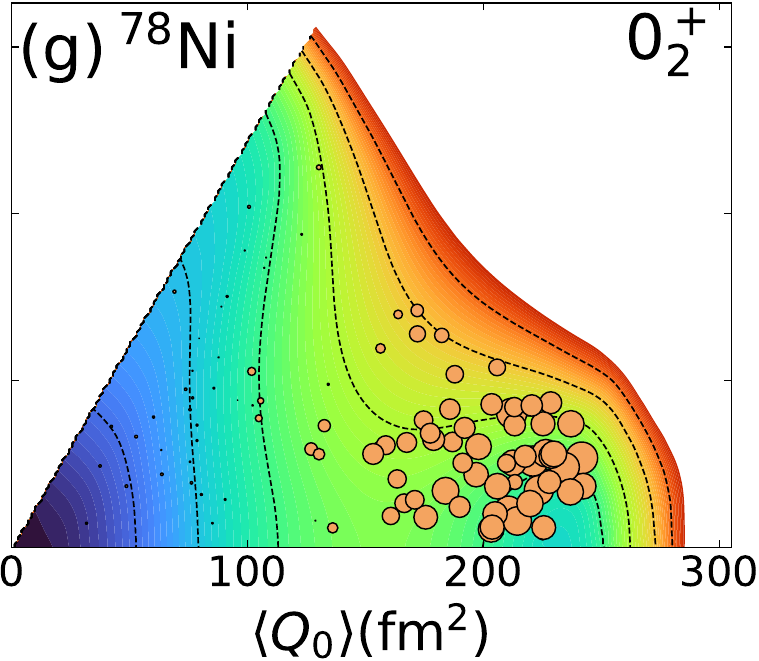}
\end{subfigure}
\hspace{-2mm}
 \begin{subfigure}{0.20\textwidth}
 \phantomcaption\label{fig:pes_n50_h}
\includegraphics[width=3.6cm, height = 3.1cm]{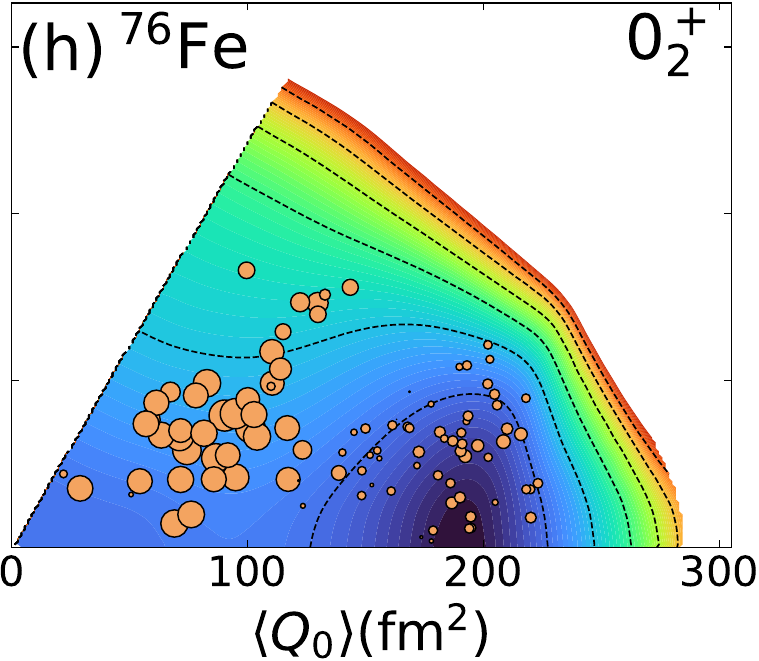}
\end{subfigure}
\hspace{-2mm}
 \begin{subfigure}{0.20\textwidth}
 \phantomcaption\label{fig:pes_n50_i}
\includegraphics[width=3.6cm, height = 3.1cm]{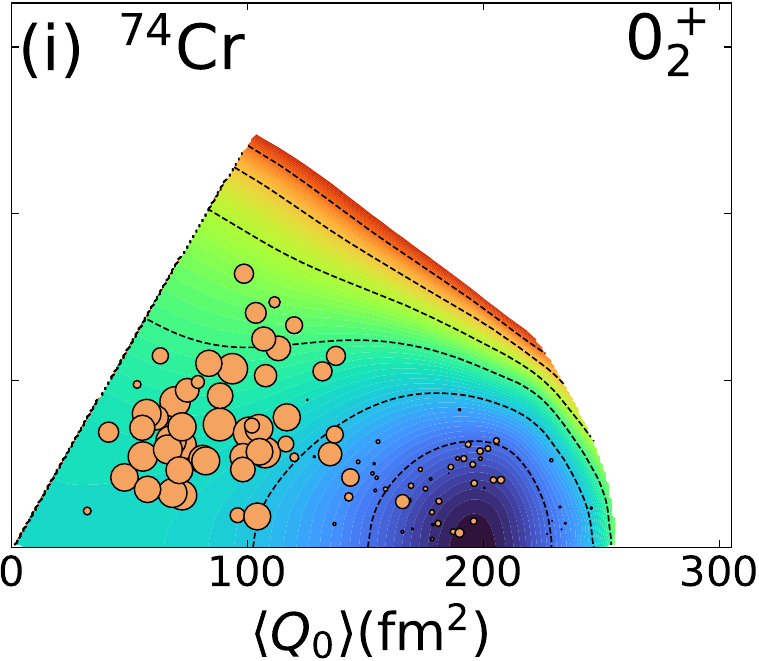}
\end{subfigure}
\hspace{-2mm}
 \begin{subfigure}{0.20\textwidth}
 \phantomcaption\label{fig:pes_n50_j}
\includegraphics[width=4.1cm, height = 3.1cm]{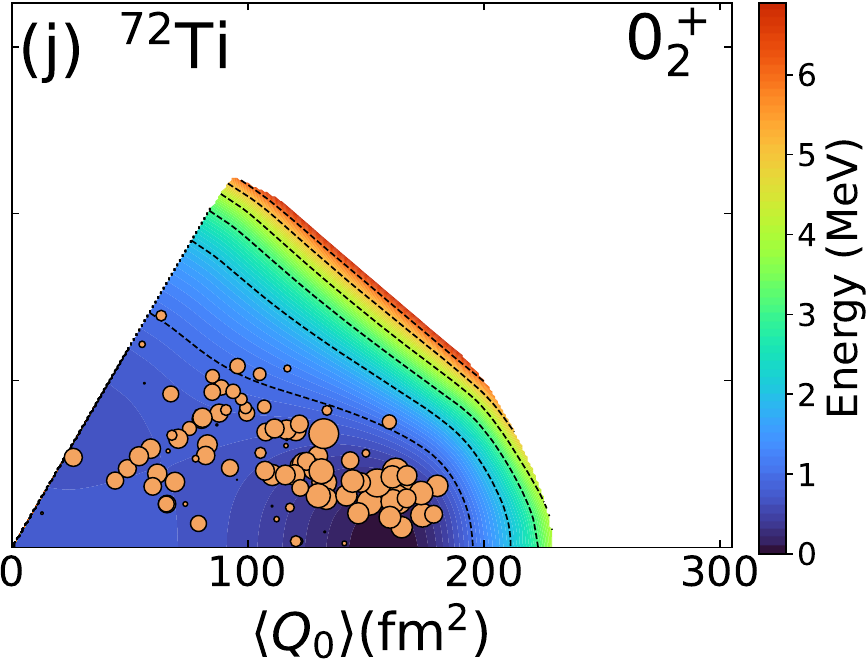}
\end{subfigure}
\caption{  Potential energy surfaces (PESs) of $N=50$ nuclei with $Z=22-30$, which is coordinated by the 
intrinsic quadrupole $Q_0$ and $Q_2$. The circles on  PESs demonstrate the shapes of the MCSM basis vectors (see the text).  
}\label{fig:pes_n50}
\end{figure*}

Further, we focus on the observed deformation in terms of the intrinsic quadrupole deformation of the MCSM wave function.
In the neutron-rich $N=50$ region, the complex interplay between normal configuration and the well-deformed intruder states is one of the most intriguing phenomena \cite{Nowacki2016PRL, yang2016PRL, Gottardo2016PRL, Nies2023PRL}, which can reveal the intrinsic structure of nuclei. 
Here, we analyze the deformation type in terms of the potential energy surfaces (PES) and MCSM basis states, "T-plot" method~\cite{Tsunoda2014RPRC}. The "T-plot" presenting the MCSM basis vectors on the PES is a crucial tool for understanding the intuitive picture of nuclei, particularly in the context of shape coexistence and shell evolutions. In this work, the PES for the present Hamiltonian within the model space is calculated using the constraint Hartree-Fock-Bogoliubov (CHFB) method, with the usual constraint on the quadrupole moments $Q_0~ (\propto\langle2z^2-x^2-y^2\rangle)$  and $Q_2~ (\propto\langle{x}^2-y^2\rangle)$ coordinates. 
Figure \ref{fig:pes_n50} displays the obtained PES for selected states of the $N=50$ isotone with $Z=22-30$  by "T-plots" coordinated by $Q_0$ and $Q_2$. We also depict scattered circles on PES corresponding to $Q_0$ and $Q_2$, which we computed using the procedure from Ref. \cite{Tsunoda2014RPRC}  for each Slater determinant $|\phi_n\rangle$. 
The circle size is proportional to the overlap probability between the resultant MCSM wave function $|\Psi\rangle$  and the normalized $P^{J\pi}_{MK}|\phi_n\rangle$ in Eq. (\ref{eqn:mcsmwave}).  Therefore, the intrinsic shape of the given state can be effectively detected by examining the size and distribution of the MCSM basis vector in the "T-plot".

The "T-plots" for the ground and the first excited $0^+$ states of $N=50$ nuclei with $Z=22-30$ in Fig. \ref{fig:pes_n50} exhibit several fundamental features. The MCSM results clearly exhibit the shape coexistence and collective behavior in $N=50$ nuclei.
In Fig. \ref{fig:pes_n50_a}, we demonstrate the PES with MCSM basis vector size for the ground $0^+_1$ state of $^{80}$Zn, where we can see that most of the large points are distributed near the spherical region. Concerning the excited $0_2^+$ state in Fig. \ref{fig:pes_n50_f}, the location of distributed points is shifted to a prolate minimum at $Q_0\sim200~{\rm fm^2}$, exhibiting large prolate deformation. The present calculation predicts the $0_2^+$ state at an excitation energy of 2.59 MeV, in reasonable agreement with the LSSM calculation using the PFSDG-U interaction, at 2.16 MeV \cite{Nies2023PRL}.  
With the decrease of two protons, 
Fig. \ref{fig:pes_n50_b} shows the PES of the $0_1^+$ state of $^{78}$Ni, where it can also be seen that most of the large circles are distributed nearly to spherical points, with a slight spreading to the oblate region. However, for the $0_2^+$ state in Fig. \ref{fig:pes_n50_g}, the locations of the distributed large circles are shifted to the pronounced prolate minimum at $Q_0\sim 200~{\rm fm}^2$, demonstrating the head of a strong-prolate-band in $^{78}$Ni. However, the $0_2^+$ has not yet been observed experimentally in $^{78}$Ni and $^{80}$Zn. It is noteworthy that the MCSM predicts the striking similarities between  $0_2^+$ states in $^{80}$Zn and the corresponding $0_2^+$ state in $^{78}$Ni,  both exhibiting a comparable deformation mechanism. However, the MCSM excitation energy of the $0_2^+$ state in $^{78}$Ni is higher than that of $^{80}$Zn due to its magicity. 

When we move below $^{78}$Ni, the ground state exhibits the more prolate-deformed minima in $^{76}$Fe, and $^{74}$Cr nuclei, where none of the demonstrated states are spherical. For $^{76}$Fe in Fig. \ref{fig:pes_n50_c}, the $0^+_1$ ground state exhibits at prolate deformed minima rather than the spherical local minimum.  
Further, a two-protons-deficient nucleus, $^{74}$Cr, demonstrates the maximum deformation for the ground $0^+_1$ state in Fig. \ref{fig:pes_n50_d}, consistent with obtaining the lowest energy of the first $2^{+}$ in Fig. \ref{fig:n50_2p_energy}. 
When we see Figs. \ref{fig:pes_n50_h}, and  \ref{fig:pes_n50_i}, the excited $0^+_2$ states of $^{76}$Fe and $^{74}$Cr exhibit the isolated behavior, becoming broad in the magnitude and $\gamma$-plane direction like triaxiality. This $0_2^+$ states of $^{76}$Fe and $^{74}$Cr exhibit the similar pattern to the ground state $0_1^+$ of $^{80}$Zn in Fig. \ref{fig:pes_n50_a}.
Figures \ref{fig:pes_n50_e} and  \ref{fig:pes_n50_j} show the same feature for $^{72}$Ti, where the local minimum in the energy landscape begins to shift back toward the spherical region.

\begin{figure}[!h]
\centering
 \begin{subfigure}{0.18\textwidth}
 \phantomcaption\label{fig:pespe_a}
\includegraphics[width = 5.0cm,height = 4.0cm]{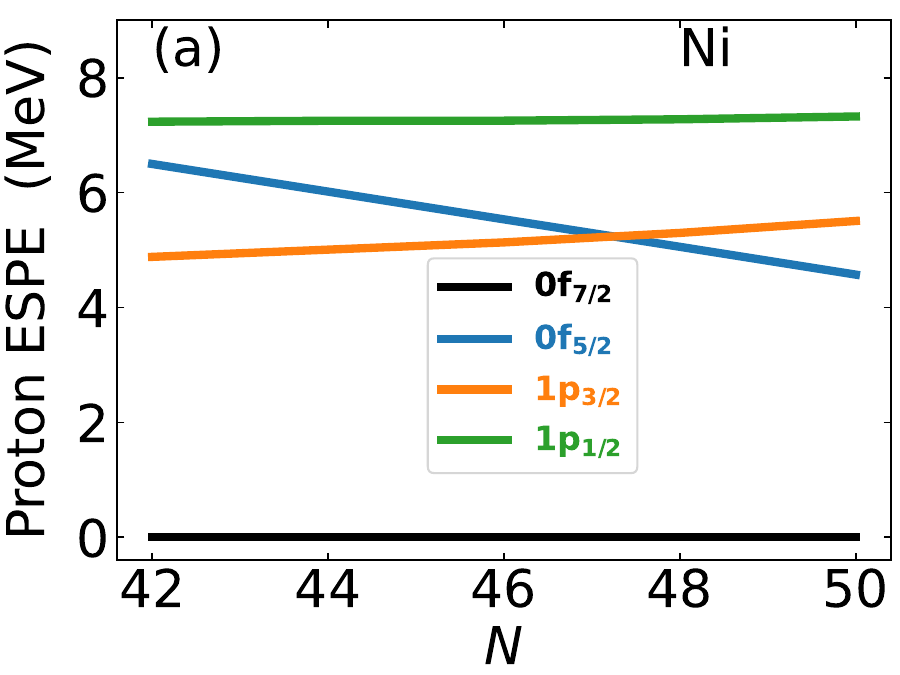}
\end{subfigure}
\hspace{1.98cm}
 \begin{subfigure}{0.12\textwidth}
 \phantomcaption\label{fig:pespe_b}
\includegraphics[width = 3.0cm,height = 4.0cm]{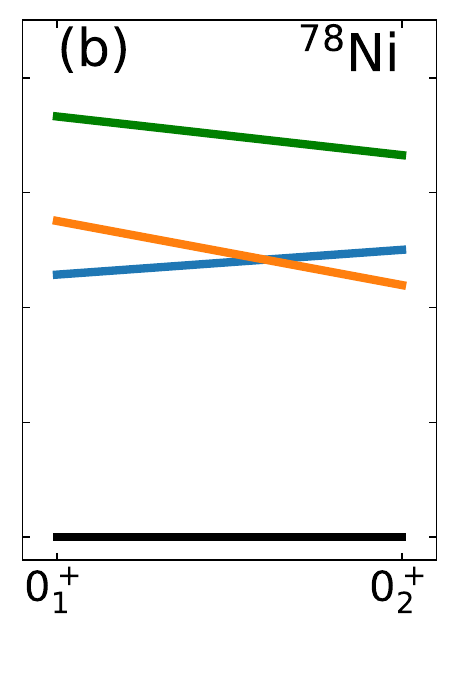}
 \end{subfigure}

\begin{subfigure}{0.18\textwidth} 
 \phantomcaption\label{fig:nespe_c}
\includegraphics[width = 5.0cm,height = 4.0cm]{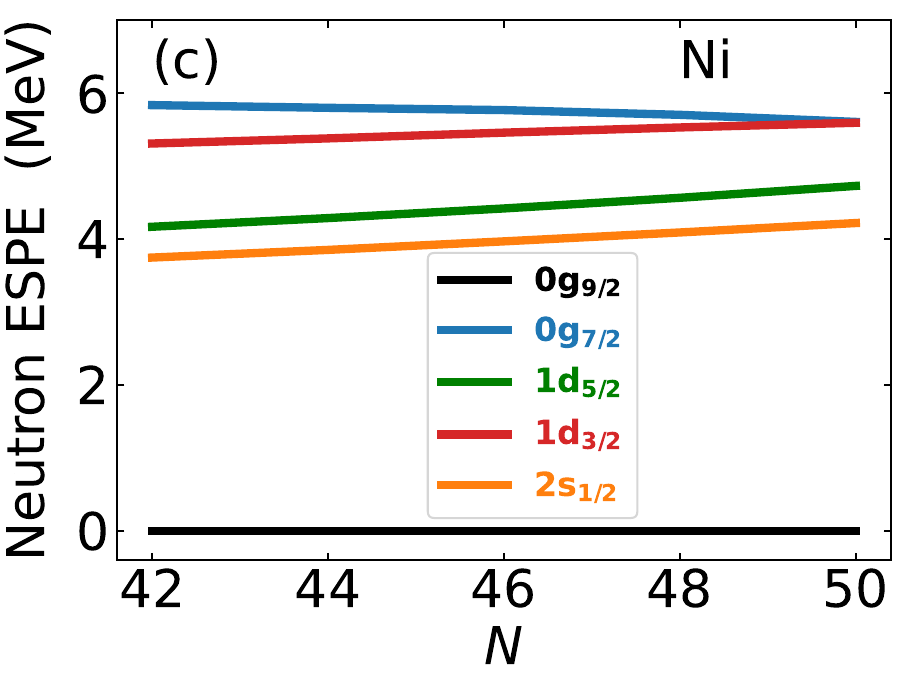}
\end{subfigure}
\hspace{1.98cm}
 \begin{subfigure}{0.12\textwidth}
 \phantomcaption\label{fig:nespe_d}
\includegraphics[width = 3.0cm,height = 4.0cm]{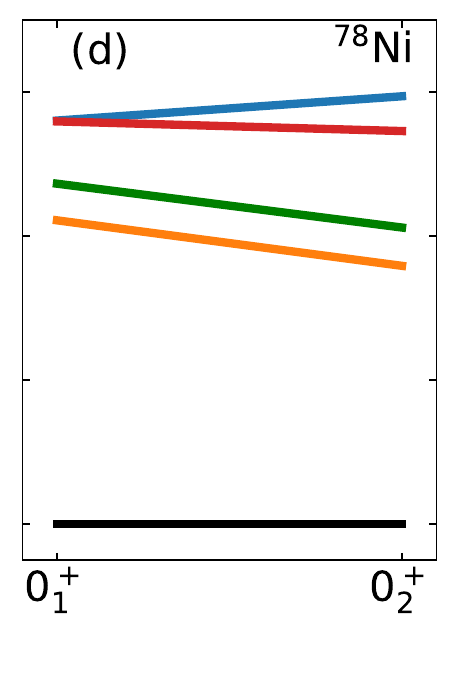}
 \end{subfigure}
\caption{ Calculated effective single-particle energies (ESPEs) for (a) the protons as a function of neutron number, (b) the $0_1^+$ ground state and the $0^+_2$ state, (c) the neutrons as a function of neutron number and (d) the ground state $0_1^+$ and first excited state $0^+_2$ using the occupation number of the resultant MCSM wave function. 
}\label{fig:espe}
\end{figure}

Deformation can also be discussed in terms of effective single-particle energies (ESPEs). Note that the ESPEs are not observable quantities, but they provide valuable insights into shell evolution. We have determined the ESPEs from the monopole component $H_m$ \cite{Otsuka2013PS, Otsuka_2016, Otsuka2020RMP} of the Hamiltonian.
In this study, we calculated the ESPEs using $\epsilon_j = \epsilon_j^0+\sum_{j'}V_{jj'}^m\langle{n_{j'}}\rangle$, where the $\epsilon_j^0$ denotes the bare SPEs, $V_{jj'}^m$ monopole component of Hamiltonian and $\la{n_{j'}}\ra$ occupation number. The occupation number $\la{n_{j'}}\ra$ are computed from the resulting MCSM wave functions.
The calculated ESPEs are shown in Fig. \ref{fig:espe}. 
Figure \ref{fig:pespe_a} demonstrates the proton ESPEs of Ni isotope as a function of the neutron number from 42 to 50. Here, we can see that the shell gap of the spin-orbit splitting of the $f_{7/2}$ and $f_{5/2}$ orbitals is reduced when neutrons occupy $g_{9/2}$ towards $N=50$. This phenomenon can be understood based on the mechanism presented in \cite{Otsuka2020RMP, Otsuka_2016}.
Since the $f_{7/2}$ and $g_{9/2}$ orbits are the upper $j_>(=l+1/2)$ type and $f_{5/2}$ orbit is of lower $j_<(=l-1/2)$ type, the proton-neutron monopole interaction of the $f_{7/2}-g_{9/2}$ ($f_{5/2}-g_{9/2}$) from the tensor force is of a repulsive (attractive) nature. Occupation of the neutron $g_{9/2}$ orbit reduces the spin-orbit splitting of proton $f_{7/2}-f_{5/2}$, which occurs by varying the neutron in $g_{9/2}$ called the {\it type I} shell evaluation.  

In the case of $^{68}$Ni, Tsunoda \textit{et al.} \cite{Tsunoda2014RPRC} claimed the deformed $0_2^+$ and $0_3^+$ states occurs due to the neutron excitation from the $fp$ shell to the $\nu{g_{9/2}}$ orbit raising the  $\pi{f_{7/2}}$ orbit and lowering the $\pi{f_{5/2}}$ orbit. Creating neutron holes in the $f_{5/2}$ orbit also 
weakens the attractive (repulsive) effects in the proton ${f_{7/2}}$ $({f_{5/2}})$ orbits. Such effects reduce the proton $f_{7/2}-f_{5/2}$ shell gap, making it $\sim 3$ MeV smaller for the $0_3^+$ state in a similar manner. This shell evolution is called {\it type II}, which is also dominated by the tensor force effect in $^{68}$Ni. 
Therefore, the smaller shell gap enhances the more particle-hole excitation, leading to stronger deformation in $^{68}$Ni.

On the other hand, for $^{78}$Ni, we find a different mechanism from the $^{68}$Ni case since the $\nu 0g_{9/2}$ orbit is occupied and the tensor force does not contribute to promoting the shape coexistence. 
 Fig. \ref{fig:pespe_b} shows the proton ESPEs of $^{78}$Ni for the ground $0_1^+$ and excited deformed $0_2^+$ states. 
The $\pi{0f_{5/2}}$ orbit is below the $\pi{1p_{3/2}}$ orbit in the $0^+_1$ state and the particle-hole excitation from $\pi{0f_{7/2}}$ to $\pi{0f_{5/2}}$ and $\pi{0p_{3/2}}$ mainly occurs in the $0_2^+$ state. Figure \ref{fig:nespe_d} shows that this excitation reduces the $N=50$ gap for the $0_2^+$ case, which causes more neutrons to be excited to the upper shell across this gap.  
The decreasing neutron occupation of the $\nu0g_{9/2}$ ($j_>'$) orbit for $0_2^+$ state increases the gap between proton spin-orbit partners by the tensor-force effect \cite{Otsuka2020RMP}, which suppresses the shape coexistence. Concerning the central force, a strong attractive force emerges between the orbital partners $n_\pi = n_\nu$ and $l_\nu\sim l_\pi+1$ with a large overlap  \cite{federman1977PL}. In the present case, the $\pi 0f_{7/2}$ orbit and the $\nu 0g_{9/2}$ orbit have a rather strong attractive force.  
Thus, by the monopole interaction of the central force, the particle-hole excitation from the $0f_{7/2}$ orbit reduces the $N=50$ gap and promotes the shape coexistence like \textrm{type II} shell evolution.

\section{Conclusions}\label{sec:conclusion}
We have constructed the shell model Hamiltonian in the $\pi{(fp)}-\nu(sdg)$ model space based on the recently developed {\it ab initio} VS-IMSRG($3{\rm f}_2$) approaches. In the present work, only the one-body part of the Hamiltonian is phenomenologically corrected minimally to reproduce the experimental data in this region. 
With the modified VS-IMSRG($3{\rm f_2}$) Hamiltonian, the advanced MCSM calculations have been performed to investigate the shell evolution, shape coexistence, and collectivity of the  $N=50$ nuclei from $^{80}$Zn towards $^{70}$Ca. With the modified Hamiltonian, the MCSM excitation spectra of $^{78}$Ni, $^{76}$Fe, $^{74}$Cr, and $^{72}$Ti largely agree with previous theoretical predictions and available experimental data. 
It enables us to discuss the prolate deformed band at high energy in $^{78}$Ni and $^{80}$Zn, which turns out to be the ground state in the  $^{76}$Fe and $^{74}$Cr nuclei. Further, we discussed the deformation band in terms of the intrinsic quadrupole of the MCSM wave function. 
In $^{68}$Ni \cite{Tsunoda2014RPRC}, the {\it type I {\rm and} II} shell evolutions are dominated by the proton-neutron tensor force. However, we predicted a different mechanism for $^{78}$Ni, where the {\it type II} shell evolution is dominated by the central force and enhances deformation. We find that the strong neutron-proton interaction between nucleons is responsible for the deformation in selected shell model orbitals. Finally, we conclude that the factorized approximation of VS-IMSRG($3{\rm f}_2$) \cite{BCHE2024PRC} provides us with a good starting point to construct a reasonable valence-space Hamiltonian, which nicely captures the deformation with minimal phenomenological adjustments to its one-body part.
This finding may suggest a future direction, i.e., the deformation could be captured by including higher-order corrections proposed in Ref.~\cite{BCHE2024PRC} only for the one-body part of the Hamiltonian, rather than the phenomenological adjustment.



\section*{Acknowledgments}
 We used the NuHamil code \cite{miyagi2023epja} and thank Ragnar  Stroberg for the IMSRG++ code \cite{imsrg++} to prepare the Hamiltonian. We would also like to thank Fr$\acute{\rm e}$d$\acute{\rm e}$ric Nowacki for the fruitful discussion.
 We acknowledge the support of the ``Program for promoting research on the supercomputer Fugaku'', MEXT, Japan (JPMXP1020230411), JST ERATO Grant No. JPMJER2304, Japan, and KAKENHI (25K00995, 25H01268, and 25K07330).
The shell-model calculations were performed using the advanced Monte Carlo Shell Model (MCSM) code \cite{shimizu2017ps} mainly on the Fugaku supercomputer at R-CCS, RIKEN (hp230207, hp240213, hp250224), and the Pegasus supercomputer at the University of Tsukuba (MCRP program: NUCLSM and wo22i002).

\bibliographystyle{apsrev4-1}
\bibliography{main}






\end{document}